\documentclass[aps,pre,reprint,groupedaddress]{revtex4-1}

\usepackage{graphicx}% Include figure files
\usepackage{amsmath}
\usepackage{bm}% bold math
\begin{document}

%\preprint{}

\title{Multifractality of complex networks}


\author{Shuhei Furuya}
 \email{sfuruya@stat.t.u-tokyo.ac.jp}
 \affiliation{Department of Mathematical Informatics, The University of Tokyo, Tokyo 113-8656, Japan}
\author{Kousuke Yakubo}%
 \email{yakubo@eng.hokudai.ac.jp}
\affiliation{Department of Applied Physics, Hokkaido
University, Sapporo 060-8628, Japan}
%\thanks{}


\date{\today}

\begin{abstract}
We demonstrate analytically and numerically the possibility
that the fractal property of a scale-free network cannot be
characterized by a unique fractal dimension and the network
takes a multifractal structure. It is found that the mass
exponents $\tau(q)$ for several deterministic, stochastic, and
real-world fractal scale-free networks are nonlinear functions
of $q$, which implies that structural measures of these
networks obey the multifractal scaling. In addition, we give
a general expression of $\tau(q)$ for some class of fractal
scale-free networks by a mean-field approximation. The multifractal
property of network structures is a consequence of large
fluctuations of local node density in scale-free networks.
\end{abstract}

 \maketitle


\section{Introduction}
Inspired by the pioneering work by Song {\it et
al.}~\cite{fractal}, the fractal property of complex networks
has been extensively studied recently
\cite{song06,goh06,song07,overlap,kim07,kawasaki,gao08,uvflower2,Mitobe2009b}.
The fractal property of a network is measured by the box-covering
method in which the minimum number of subgraphs (boxes) of diameter
$l$ (in the sense of network distance) required to cover the fractal
network is proportional to $l^{-D_{\text{f}}}$ with the fractal
dimension $D_{\text{f}}$. Most of real-world fractal networks
are inhomogeneous in the sense of the scale-free property
defined by a power-law degree distribution $P(k) \propto
k^{-\gamma}$, where $k$ is the number of connections of a node
(degree) \cite{scale-free}. Thus, the number of nodes in a
subgraph of size $l$ depends strongly on whether the subgraph
includes hubs and their neighboring nodes or not, which implies
that the distribution of local node density is highly
inhomogeneous. An inhomogeneous distribution of a physical
quantity on a fractal object often exhibits the multifractal
property \cite{paladin87,multifractal_stanley,mandelbrot74,witten81}. 
In many of fractal objects embedded in Euclidean space, however,
the underlying structure seldom shows the multifractal nature
because of an exponentially thin tail of the mass distribution.
On the contrary, we expect the multifractal scaling in a
scale-free network due to large fluctuations of local node density.
In this paper, we show analytically and numerically that fractal
scale-free networks (FSFNs) can have the multifractal property
in their structural features.

\section{Multifractal analysis of networks}
In order to explicate the possibility that a FSFN takes a
multifractal structure, let us consider, at first, why
conventional fractal structures do not possess the multifractal
property. In the multifractal analysis, the behavior of a
coarse-grained physical quantity on a fractal object is argued.
Let $x_i$ and $\mu_{i}$ be a physical quantity on the
discretized position $i$ and its normalized value (measure),
this is,
\begin{equation}
\label{measure1}
\mu_{i}=\frac{x_{i}}{\sum_{j}x_{j}} \ .
\end{equation}
The coarse-grained measure (box measure) $\mu_{b(l)}$ is then
given by $\mu_{b(l)}=\sum_{i\in b(l)} \mu_{i}$, where $b(l)$ is
a box of size $l$ in the system. If the spatial distribution of
the physical quantity $x$ does not bring a characteristic
length scale, the $q$th moment of the box measure
\begin{equation}
\label{box_measure}
\langle \mu_{l}^{q}\rangle=\sum_{b(l)} \mu_{b(l)}^{q}
\end{equation}
has a power-law $l$ dependence, namely, $\langle
\mu_{l}^{q}\rangle\propto l^{\tau(q)}$, where the summation in
Eq.~(\ref{box_measure}) is taken over boxes of size $l$
required to cover minimally the entire system. In the case that
the exponent $\tau(q)$ (called as the mass exponent) is a
nonlinear (linear) function of $q$, the distribution of the
measure is regarded to be multifractal (unifractal). Here, we
consider a constant mass of the site $i$ as a physical quantity
$x_{i}$. The normalized measure $\mu_{i}$ representing the mass
density becomes constant and the box measure $\langle
\mu_{b(l)} \rangle$ averaged over boxes is proportional to
$l^{D_{\text{f}}}$, where $D_{\text{f}}$ is the fractal
dimension of the system. If the fluctuation of $\mu_{b(l)}$
over boxes is sufficiently small, Eq.~(\ref{box_measure}) is
approximated as
\begin{equation}
\langle \mu_{l}^{q}\rangle\sim \sum_{b(l)} \langle \mu_{b(l)}\rangle^{q}
\sim l^{D_{\text{f}}(q-1)}\ .
\label{similar}
\end{equation}
Therefore we have the linear relation
$\tau(q)=D_{\text{f}}(q-1)$ representing the unifractal nature
of the mass density distribution. In fact, a narrow
distribution of the box measure is widely observed in many
fractals embedded in Euclidean space
\cite{Bramwell1998,Mitobe2009a,zheng03,binder81,federrath10}.
Thus, most of fractal systems take unifractal structures, with
some exceptions such as mathematical multifractal sets as the
two-scale Cantor set \cite{Ficker1989} and geochemical
distribution of minerals
\cite{Goncalves2001,Agterberg2007,Xie2010}. The box measure
$\mu_{b(l)}$ of node density in a FSFN, however, has large
fluctuations over boxes \cite{Mitobe2009b}. This is because
$\mu_{b(l)}$ of a box including a hub node can be much larger
than that of a box without hubs. If the probability
distribution function of the box measure has a fat tail like a
power-law or log-normal form, $\mu_{b(l)}$ cannot be
approximated by $\langle\mu_{b(l)}\rangle$ and the failure of
Eq.~(\ref{similar}) gives a possibility of the multifractal
scaling in structures of complex networks.

When applying the multifractal analysis to complex networks, it
is unavoidable that covering boxes (subgraphs) of network
diameter $l$ overlap each other \cite{overlap}, if every box
includes the maximum number of nodes unless the subgraph
diameter does not exceed $l$ as in the conventional
multifractal analysis \cite{paladin87}. Therefore, the measure
$\mu_{i}$ defined by Eq.~(\ref{measure1}) cannot be normalized as
$\sum_{b(l)}\mu_{b(l)}=1$, and the mass exponent does not
satisfy the basic condition $\tau(1)=0$. In order to overcome
this difficulty, we modify the definition of the measure as
\begin{equation}
\mu_i = \frac{x_i}{\sum_{b(l)} \sum_{j \in b(l)} x_j}.
\label{mu_CN}
\end{equation}
In this definition, the normalization constant, and then
$\mu_{i}$, varies with the box size $l$. It is easy to confirm
that the mass exponent calculated from Eq.~(\ref{mu_CN})
satisfies the general conditions $\tau(0)=-D_{\text{f}}$ and
$\tau(1)=0$ and for a unifractal system
$\tau(q)=D_{\text{f}}(q-1)$. We set hereafter $x_{i}=1$ to
analyze the multifractal nature of the node density in a
complex network.
\begin{figure}[ttttt]
\begin{center}
\includegraphics[width=0.48\textwidth]{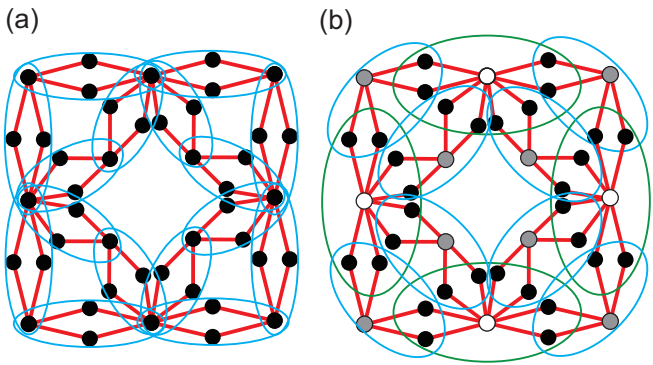}
\caption{(Color online) Two types of covering schemes for the
$(u,v)$ flower with $u=v=2$ in the third generation. (a) The
network is covered by $16$ $(u,v)$ flowers in the first
generation (scheme I), and (b) the network is first covered by
four subgraphs of size $l=2$ centered at the largest hubs (white
nodes) then covered by eight subgraphs centered at the second
largest hubs (gray nodes) (scheme II).}
\label{fig1}
\end{center}
\end{figure}
\section{Multifractality of $(u,v)$ flowers}
Using the above definition of $\mu_{i}$, we first examine
analytically the multifractal property of the $(u,v)$-flower
model \cite{uvflower2}. This deterministic model provides a
class of FSFNs. In this model, we start with the cycle graph
consisting of $w\equiv u+v$ ($1<u\le v$) nodes and edges (the
first generation). The network in the $n$th generation is
obtained by replacing each edge in the $(n-1)$th generation
network by two parallel paths of $u$ and $v$ edges. The network
with large $n$ has the scale-free property with the degree
exponent $\gamma=1+\log w/\log 2$ \cite{uvflower2}. The number
of nodes in the $n$th generation network is given by
\begin{equation}
\nu_n=\left(\frac{w-2}{w-1}\right)w^n+\left(\frac{w}{w-1}\right),
\label{size}
\end{equation}
and the number of edges is $w^n$. In addition, when $w$ is
even, the diameter is written as
\begin{equation}
L_n=\left(\frac{u+v}{2}+\frac{v-u}{u-1} \right)u^{n-1}-\frac{v-u}{u-1}\ .
\label{diameter}
\end{equation}
From the network formation algorithm, the $n$th generation
network is constructed by $w^{n-m}$ of $(u,v)$ flowers in the
$m$th generation, where $m<n$. This implies that if we cover
the $n$th generation network by the $m$th generation subgraphs
as shown in Fig.~1(a) the number of covering boxes
$N_{b(L_{m})}$ is $w^{n-m}$. Using Eq.~(\ref{diameter}), we can
rewrite this relation as
$N_{b(L_{m})}=w^{n}[(L_{m}+b)/a]^{-\log w/\log u}$, where
$a=(w/2+b)/u$ and $b=(v-u)/(u-1)$. If the length $L_{m}$ is
large enough, we have the fractal dimension given by
$D_{\text{f}}=\log w/\log u$ \cite{uvflower2}. The above
covering scheme (named as the covering scheme I), however, does
not lead the minimum value of $N_{b(L_{m})}$ for a fixed
$L_{m}$. Let us consider the following covering scheme (named
as the covering scheme II). At first, we cover the
$(u,v)$ flower by subgraphs of size $L_{m}$ centered at the
largest hubs, then the remaining network is covered by
subgraphs centered at the second largest hubs. Repeating this
procedure until all nodes are covered by subgraphs [as shown in
Fig.~1(b)] the number of covering boxes $N_{b(L_{m})}$ becomes
less than $w^{n-m}$. Although two covering schemes yield
different $N_{b(L_{m})}$, these are in proportion as shown
later, and then the fractal dimension calculated by the
covering scheme II is the same as that by scheme I.
Therefore, both covering schemes are valid for the
\textit{fractal} analysis. In the \textit{multifractal}
analysis, however, we treat not only $N_{b(L_{m})}(=\langle
\mu_{L_{m}}^{q=0}\rangle)$ but also $\langle
\mu_{L_{m}}^{q}\rangle$ for any $q$. Since the moment $\langle
\mu_{L_{m}}^{q}\rangle$ calculated by the covering scheme I is
generally not proportional to $\langle \mu_{L_{m}}^{q}\rangle$
by scheme II, we need to choose scheme II with less
(probably minimum) covering boxes for the multifractal analysis
of the $(u,v)$ flower. It is, in general, quite important to
cover a network by less number of boxes in the multifractal
analysis comparing to the case of the fractal analysis.

Let us cover the $(u,v)$ flower in the $n$th generation by
boxes of size $l=L_{m}$ ($1\ll m\ll n$) in the covering scheme
II. The number of boxes $N_{b(s,L_{m})}$ centered at the $s$th
largest hubs is equal to the number of such hubs, thus we have
\begin{equation}
\label{num_box}
N_{b(s,L_{m})}=\nu_{s}-\nu_{s-1} \qquad (1\le s\le n-m)\ ,
\end{equation}
where $b(s,L_{m})$ represents a box of size $L_{m}$ centered at
the $s$th largest hub and $\nu_{0}=0$. Since the number of
nodes $\tilde{\nu}_{s}(L_{m})$ in $b(s,L_{m})$ is presented by
\begin{equation}
\label{num_node}
\tilde{\nu}_{s}(L_{m})=2^{n-m-s+1}(\nu_{m}/2+1)\simeq 2^{n-m-s}\nu_{m}\ ,
\end{equation}
the total number of nodes in all boxes, namely, the denominator
of Eq.~(\ref{mu_CN}), is given by
$\sum_{s=1}^{n-m}N_{b(s,L_{m})} \tilde{\nu}_{s}(L_{m})$. Using
Eqs.~(\ref{size}), (\ref{num_box}), and (\ref{num_node}), the
measure $\mu_{i}$ defined by Eq.~(\ref{mu_CN}) with $x_{i}=1$
is thus expressed as
\begin{equation}
\label{measure_uv}
\mu_{i}=\frac{w-1}{w^{n}(w-2)}\ .
\end{equation}
It should be noted that $\mu_{i}$ for the $(u,v)$ flower is
independent of the box size though the box measure given by
Eq.~(\ref{mu_CN}) generally depends on $l$. Denoting $\mu_{i}$
independent of $i$ by $\mu$, the $q$th moment of the box
measure is calculated from $\langle
\mu_{L_{m}}^{q}\rangle=\sum_{b(s,L_{m})} \left(\sum_{i\in
b(s,L_{m})}\mu\right)^{q}= \sum_{s=1}^{n-m}
N_{b(s,L_{m})}[\mu\tilde{\nu}_{s}(L_{m})]^{q}$. By means of
Eqs.~(\ref{size}) and (\ref{num_box})$-$(\ref{measure_uv}), the
quantity $\langle \mu_{L_{m}}^{q}\rangle$ is calculated as
\begin{multline}
\langle \mu_{L_{m}}^{q} \rangle = \frac{w^{n(1-q)}}{W_{q}}
\Bigg[\left(\frac{2^{q}}{w}\right)^{n-1}(W_{q}-1)
  \left(\frac{L_{m}+b}{a} \right)^{\frac{q\log (w/2)}{\log u}} \\
+ \left(\frac{L_{m}+b}{a} \right)^{\frac{(q-1)\log w}{\log u}}
\Bigg]\ , \label{moment_uv}
\end{multline}
where $a$ and $b$ are defined below Eq.~(\ref{diameter}),
$W_{q}=(w-2^{q})/(w-2)$, and the approximation $\nu_{m}\simeq
w^{m}(w-2)/(w-1)$ is used. The dominant term in
Eq.~(\ref{moment_uv}) for large $L_{m}$ depends on $q$, and the
mass exponent is given by
\begin{equation}
\tau (q)=
\begin{cases}
\displaystyle q\frac{\log{(w/2)}}{\log{u}} & \text{if } \displaystyle q \ge \frac{\log{w}}{\log{2}}  \vspace{2mm},\\
\displaystyle (q-1)\frac{\log{w}}{\log{u}} & \text{if } \displaystyle q < \frac{\log{w}}{\log{2}}.
\end{cases}
\label{tauq}
\end{equation}
The nonlinear form of $\tau(q)$ indicates that local node
densities of the $(u,v)$ flower are distributed in a
multifractal manner \cite{footnote}. Although the above
argument holds only for even values of $w$, we found that
Eq.~(\ref{tauq}) is a good approximation also for odd $w$. It
should be noted that $\langle \mu_{L_{m}}^{q}\rangle
=w^{n(1-q)}[(L_{m}+b)/a]^{(q-1)\log w/\log u}$ obtained by
scheme I is proportional to Eq.~(\ref{moment_uv}) for $q<\log
w/\log 2$, but not otherwise.

It is generally difficult to find the way to cover minimally a
given network because the minimization of the number of
covering boxes is known to be NP hard. We then need to cover the
network by a \textit{less} number of boxes, as an
approximation, in actual multifractal analyses. A variety of
such covering methods have been proposed so far
\cite{fractal,overlap,goh06,song07,kim07,gao08,kawasaki}.
In order to clarify whether such covering techniques proven to
be efficient for fractal analyses still work even in
multifractal analyses sensitive to the covering way, we compare
the analytical expression Eq.~(\ref{tauq}) with the numerically
calculated $\tau(q)$ by adopting the compact-box-burning
algorithm \cite{song07} modified to shorten the computing time
\cite{kawasaki}. The results shown in Fig.~2(a) clearly
demonstrate the validity of this covering method.

\begin{figure}[ttttt]
\begin{center}
\includegraphics[width=0.48\textwidth]{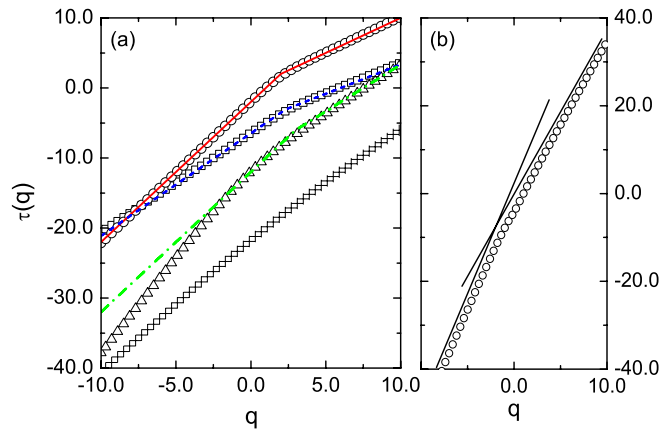}
\caption{(Color online) Mass exponent $\tau(q)$ for several
fractal complex networks. (a) Solid line (red line) and
circles indicate $\tau(q)$ for the $(u,v)$ flower with $u=v=2$
calculated by Eq.~(\ref{tauq}) and by the numerical
box-covering for the network in the eighth generation
($N=43,692$), respectively. Squares represent numerical results
of $\tau(q)$ for the network formed by the minimal model with
$m=2$ in the seventh generation ($N=15,626$). Dashed line (blue
line) indicates the theoretical curve for this minimal
model predicted by Eq.~(\ref{mean_field}). Triangles and
crosses show $\tau(q)$ for the giant components of the fitness
model ($N=100,000$) and the Erd\H{o}s-R\'enyi random graph
($N=200,000$) at their percolation transitions, respectively.
The degree exponent for the fitness model is set to be
$\gamma=4.0$. We averaged $\langle \mu_{l}^{q}\rangle$ over
$100$ realizations both for the fitness and the random graph
models. Dashed-dotted line (green line) indicates
$\tau(q)$ given by Eq.~(\ref{mean_field}) with $\gamma=4.0$ and
$D_{\text{f}}=2$ corresponding to the fitness model. Results
except for the $(u,v)$ flower are shifted vertically for
clarity though all $\tau(q)$ actually pass through
$\tau(1)=0$. (b) The mass exponent $\tau(q)$ for the WWW is
presented by circles. Straight lines indicating the slopes of
$\tau(q)$ for $q\ll -1$ and $q\gg 1$ are just guides to the
eye.}
\label{fig2}
\end{center}
\end{figure}
\section{Mean-field argument}
Let us generalize our argument to some extent. We treat a FSFN
of $N$ nodes, whose degree distribution is given by $P(k)\propto k^{-\gamma}$.
Covering the network minimally by $N_{b(l)}$ boxes of diameter $l$,
the mean number of nodes $\langle \nu_{l}\rangle$ in a box is
given by $N/N_{b(l)}$, where $N_{b(l)}\propto l^{-D_{\text{f}}}$.
Regarding mutually connected boxes of size $l$ as a renormalized
network \cite{fractal}, the fractal property of the original network assures the
relation $P_{l}(k_{l})\propto k_{l}^{-\gamma}$, where $P_{l}(k_{l})$
is the degree distribution function of the renormalized network.
Here, we assume that each renormalized node is statistically equivalent
and the number of nodes $\nu_{l}(k_{l})$ in a covering box corresponding
to the renormalized node of degree $k_{l}$ has negligibly small fluctuations
over boxes. In this mean-field approach, the quantity $\nu_{l}(k_{l})$
is represented by its mean value,
\begin{equation}
\label{assamption}
\nu_{l}(k_{l})=\frac{k_{l}}{\langle k\rangle}\langle \nu_{l}\rangle\ ,
\end{equation}
where $\langle k\rangle=\langle k_{l}\rangle$ is the average degree
of the original network. The $(u,v)$ flower satisfies Eq.~(\ref{assamption})
rigorously in the thermodynamic limit ($n\to \infty$). Since
the box measure $\mu_{b(l)}$ is given by $\nu_{l}(k_{l})$ normalized
by $N$, this is, $\mu_{b(l)}=k_{l}/[\langle k\rangle
N_{b(l)}]\propto (k_{l}/\langle k\rangle)l^{D_{\text{f}}}$, the
$q$th moment $\langle \mu_{l}^{q}\rangle$ can be calculated by
Eq.~(\ref{box_measure}). Considering that the maximum degree
$k_{\text{max}}$ is proportional to $N^{1/(\gamma-1)}$, we have
$\langle \mu_{l}^{q}\rangle\propto l^{(q-1)D_{\text{f}}}$ for
$q<\gamma-1$ and $\langle \mu_{l}^{q}\rangle\propto
l^{qD_{\text{f}}(\gamma-2)/(\gamma-1)}$ for $q\ge \gamma-1$.
Therefore, the mass exponent $\tau(q)$ of this network is
presented by
\begin{equation}
\label{mean_field}
\tau(q)=
\begin{cases}
(q-1)D_{\text{f}} & \text{if $q<\gamma-1$} \\
qD_{\text{f}}\left(\dfrac{\gamma-2}{\gamma-1}\right) & \text{if $q\ge \gamma-1$} \ .
\end{cases}
\end{equation}
This result implies that FSFNs satisfying Eq.~(\ref{assamption})
generally take multifractal (bifractal) structures. Our result
Eq.~(\ref{tauq}) for the $(u,v)$ flower is a special case of
Eq.~(\ref{mean_field}).

In order to confirm the validity of the above mean-field
argument, we numerically calculate the mass exponent for the
minimal model proposed by \cite{song06}, employing the
compact-box-burning algorithm \cite{song07}. A network formed
by this model possesses the scale-free property with the degree
exponent $\gamma=1+\log(2m+1)/\log m$ and takes a fractal
structure with the fractal dimension $D_{\text{f}}=\log(2m+1)/\log 3$.
The minimal model also satisfies Eq.~(\ref{assamption}) as in the case of
the $(u,v)$ flower model \cite{song06}. The nonlinear behavior of numerically calculated
$\tau(q)$ indicated by squares in Fig.~2(a) agrees quite well
with the theoretical prediction Eq.~(\ref{mean_field}). As an
example of networks not satisfying Eq.~(\ref{assamption}) due to
large fluctuations of $\nu_{l}(k_{l})$, we treat a stochastic model of
FSFNs, namely the fitness model proposed by
\cite{caldarelli02}. A network formed by this algorithm also
has the scale-free property and the giant component exhibits
the fractal nature at the percolation transition
\cite{kawasaki}. Triangles in Fig.~2(a) represent $\tau(q)$ for
FSFNs formed by the fitness model with $\gamma=4.0$ and
$D_{\text{f}}=2$ \cite{kawasaki}. The exponent $\tau(q)$ is
also a nonlinear function of $q$, which suggests the
multifractal structure of the network, but cannot be described
by Eq.~(\ref{mean_field}). Finally, we calculate the mass
exponent $\tau(q)$ for the World Wide Web (WWW) with $325,729$
nodes \cite{WWW}, which is known to be a representative
real-world FSFN \cite{fractal}. Although the nonlinearity of $\tau(q)$ is
weak as shown in Fig.~2(b) and the result is not described by
Eq.~(\ref{mean_field}), two tangential lines in the extreme
regimes $q\ll -1$ and $q\gg 1$ have definitely different
slopes, which shows the multifractal structure of the WWW. The
multifractal property found in these networks is obviously
caused by the scale-free nature of networks. In fact, $\tau(q)$
for the giant component of the Erd\H{o}s-R\'enyi random graph
\cite{erdos59} at the percolation threshold is, as shown by
crosses in Fig.~2(a), a linear function of $q$, where the giant
component is fractal but not scale free.

\section{Summary}
In conclusion, we demonstrated analytically and numerically
that fractal scale-free networks (FSFNs) can take multifractal
structures in which the fractal dimension is not enough to
characterize fractality of systems. The multifractal nature is
caused by large fluctuations in local node density in
scale-free networks. Although all examples treated in this
paper exhibit the multifractal nature, further investigations
are needed to clarify whether any FSFNs take multifractal
structures. It is also crucial to study how the multifractal
property affects physical phenomena or dynamics on complex
networks.

\begin{acknowledgements}
This work was supported by a Grant-in-Aid
for Scientific Research (No.~22560058) and Grant-in-Aid for
JSPS Fellows (No.~22$\cdot$3380) from Japan Society for the Promotion
of Science. Numerical calculations in this work were performed
on the facilities of the Supercomputer Center, Institute for
Solid State Physics, University of Tokyo.
We are grateful to a referee for
the crucial idea for the mean-field argument that improves
the paper drastically.
\end{acknowledgements}

%\begin{acknowledgments}
%\end{acknowledgments}

%\bibliography{furuya_ref.bib}

\end{document}